\documentclass[twocolumn,amssymb,prb,superscriptaddress,floats,showkeys,showpacs]{revtex4-1}

\usepackage[T1]{fontenc}
\usepackage[latin9]{inputenc}
\usepackage{booktabs}
\usepackage{graphicx}

\usepackage{bm}
\usepackage{textcomp}
\usepackage{epstopdf}
\usepackage{braket}

\begin{document}

\title{Atomistic theory of dark excitons in self-assembled quantum dots \\
of reduced symmetry}

\author{M. Zieli\'{n}ski}

\email{mzielin@fizyka.umk.pl}

\affiliation{Institute of Physics, Faculty of Physics, Astronomy and Informatics,
Nicolaus Copernicus University, Grudzi\k{a}dzka 5, 87-100 Toru\'{n}, Poland}

\author{Y. Don}

\author{D. Gershoni}

\affiliation{The Physics Department and the Solid State Institute, Technion--Israel
Institute of Technology, Haifa, 32000, Israel}

\begin{abstract}
	We use an atomistic model to consider the effect of
	shape symmetry breaking on the optical properties of self-assembled
	InAs/GaAs quantum dots. In particular, we investigate the energy level
	structure and optical activity of the lowest energy excitons in these
	nanostructures.  We compare between quantum dots with two-fold rotational
	and two reflections ($C_{2v}$) symmetry and quantum dots in which this
	symmetry was reduced to one reflection only ($C_{s}$) by introducing a facet
	between the quantum dots and the host material. We show that the symmetry
	reduction mostly affects the optical activity of the dark exciton. While in
	symmetric quantum dots, one of the dark exciton eigenstates has a small dipole
	moment polarized along the symmetry axis (growth direction) of the quantum
	dot, in non-symmetric ones,
	%the two dark excitons' dipole moments are very uneven and predominantly
	%cross linearly polarized perpendicularly to growth direction.
	the two dark excitons' dipole moments are predominantly cross-linearly
	polarized perpendicular to the growth direction and reveal pronounced
	polarization anisotropy. Our model calculations agree quantitatively with
	recently obtained experimental data.
\end{abstract}



\maketitle

%%%%%%%%%%%%%%%%%%%%%%%%%%%%%% Bulk text

\section{Introduction}

Confined excitons in single semiconductor quantum dots (QDs) have
been a subject of extensive studies since they play a central role
in many schemes which lead towards applications in quantum optics
and future quantum technologies.~\cite{loss} If a QD maintains
two-fold rotational symmetry (such as $C_{2v}$), its lowest-energy
excitonic eigenstates can be divided into two characteristic doublets:
the lowest energy doublet which is mostly optically inactive --- the
dark exciton (DE) --- and the higher energy doublet which constitutes
the fundamental optical excitations of the QD --- the bright exciton.~\cite{ivchenko,bayer}
Bright excitons (BEs) have been thoroughly studied both experimentally
and theoretically due to their obvious use in applications based on
single photon sources,~\cite{single-photon-sources,Santori,yuan2002}
single photon detectors,~\cite{single-photon-detectors1,single-photon-detectors2,single-photon-detectors3}
entangled photon sources~\cite{entangled-photon1,entangled-photon2,muller2014}
and photon-spin entanglement.~\cite{photonspin1,photonspin2,photonspin3}
On the contrary, relatively little is known about the nature of the
dark excitons since their optical inactivity renders them quite difficult
to access experimentally. Recently, however, it was demonstrated that
QD-confined DEs, despite their weak optical activity in emission,
can be efficiently accessed by optical absorption and by charge injection.~\cite{poem2010,dark-experiment}
In this way, it was demonstrated that the DE actually forms a long
lived two-level system (qubit)~\cite{poem2010} with a long coherence
time.~\cite{dark-experiment} As a neutral integer spin qubit, the
DE has some obvious advantages~\cite{dark-experiment} over the
more conventional single carrier spin-based qubits.~\cite{kim2010,press2010,ramsay2008-2,greilich2011,degreve2011,godden2012}
For these and other reasons, there is an increasing scientific interest
in DE studies.~\cite{poem2010,dark-smolenski,dark-experiment}

From general theoretical considerations based on group theory, one
expects that in symmetric QDs, the BE eigenstates have cross-linearly
in-plane polarized dipole moments ($\hat{x}$ and $\hat{y}$ directions,
respectively), one of the DE eigenstates is completely dark (\emph{i.e.}~its
dipole moment vanishes), and the other DE eigenstate may have a
non-vanishing dipole moment, polarized along the QD symmetry axis
($\hat{z}$ direction).~\cite{karlsson2,ivchenko,karlsson1} In
order to obtain quantitative information about the magnitude of these
dipole moments, however, more detailed theoretical modeling based
on either effective mass approximations~\cite{ivchenko,takagahara,poem2007}
or atomistic calculations~\cite{karlsson1,karlsson2,garnett-prl,zielinski-prb2010,dark-marek}
are required. Comparison with experimental studies~\cite{dark-smolenski,bayer,alon2006,gammon,dark-experiment}
are at times not easily obtained due to the lack of information regarding
the actual size, composition, structure, and strain of an experimentally
measured QD. In addition, the usually very weak optical activity of
the DE is yet another obstacle to obtaining accurate experimental
data. This obstacle can be partially removed by applying an external
magnetic field perpendicular to the QD symmetry axis (Voigt geometry).
Such a perturbation mixes between the BE and DE eigenstates, such
that the mixed DE eigenstate gains a dipole moment and becomes optically
active with in-plane polarized dipole moments, like those of the
BE eigenstates. Extrapolations to zero magnetic field may yield information
about the bare DE eigenstates as well.~\cite{dark-smolenski,bayer,alon2006,gammon,dark-experiment}

Modeling and structural simulations show that the magnitude of the
dipole moment of the DE strongly depends on the actual shape of the
QD and on its symmetry. For symmetric InAs/GaAs self-assembled QDs,
it is generally accepted that the $\hat{z}$-polarized optical activity
of the DE is due to hole subband mixing.~\cite{ivchenko,karlsson2,dark-smolenski,dark-marek,zielenski2013}
Previous experimental studies which reveal residual unexplained in-plane
optical activity of dark excitons attributed it to the reduced symmetry
(lower than $C_{2v}$) of the QD.\cite{bayer} Bayer \emph{et~al.}~\cite{bayer}
suggested that the reduction in symmetry, like the in-plane external
magnetic field, mixes between the BE and DE eigenstates, causing the
later to become optically active.

In this paper, we used an atomistic model to realistically consider
confined excitons in InGaAs/GaAs self-assembled semiconductor QDs
of various shapes which maintain $C_{2v}$ symmetry. We then introduce
a structural perturbation in the QD shape which reduces the symmetry
to $C_{s}$. We show indeed that such a perturbation is sufficient
to trigger in-plane polarized optical activity of the DE eigenstates.
We establish this way that while the in-plane (out-of-plane polarized)
emission from the DE eigenstates can indeed be attributed to hole
subband mixing, the out-of-plane (in-plane polarized) DE dipole moment
is a very sensitive probe for the QD deviation from symmetry, and
it mainly results from DE-BE mixing.

The use of an atomistic model is essential for an accurate, realistic
description of the QD structure while maintaining the crystal lattice
symmetry and keeping atomic scale details of the composition and strain
distribution. We utilize empirical tight-binding theory for single
particle states with an $sp^{3}d^{5}s^{*}$ orbitals,~\cite{jancu}
nearest-neighbor coupling, and spin-orbit and strain effects included.~\cite{zielinski-including}
Relaxation of strain is accounted for via atomistic valence force field
theory.~\cite{keating,martin} Piezoelectricity is included as well using
both first- and second-order terms.~\cite{bester-piezo} Exciton
states are calculated using configuration-interaction treatment,
which includes all possible determinants constructed from the 12 lowest-energy
electron and 12 lowest-energy hole states (including spin).~\cite{sheng-cheng-prb2005}
Finally, the optical spectrum is obtained by calculating the photoluminescence
(PL) intensity due to the recombination of one electron-hole pair
in a given exciton state using Fermi's golden rule.~\cite{zielinski-prb2010}

\section{High-symmetry quantum dots}

\begin{figure}
	\begin{centering}
		\includegraphics[width=0.5\textwidth]{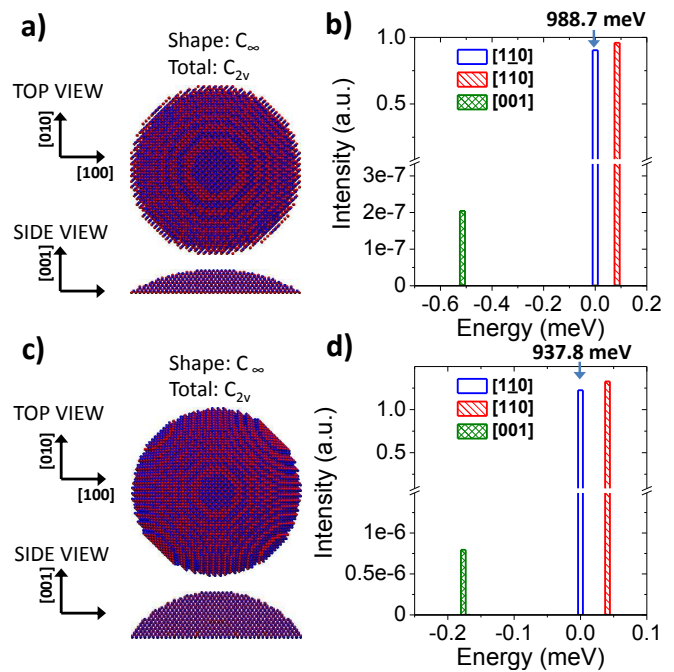}
		\protect\caption{ (Color online) Schematic description of InAs/GaAs $C_{2v}$ lens-shaped
			QD of 18 nm circular base radius and 2.4 nm (a) and 4.8 nm (c) height.
			Indium (Arsenide) atoms are shown as blue (red) spheres. Wetting layer
			atoms, as well as surrounding GaAs matrix atoms, are not shown. The
			calculated excitonic spectrum for the QDs in (a) and (c) are displayed
			in (b) and (d) respectively.}
		\label{lens}
		\par
	\end{centering}
\end{figure}

Fig.~\ref{lens} (a) and (c) show the geometry of two lens shaped QDs with the
same circular base radius ($18$~nm) but different heights: $2.4$~nm and
$4.8$~nm, respectively. In Fig.~\ref{lens} (b) and (d) we present the
calculated excitonic emission spectra for these two symmetrical InAs/GaAs QDs.
%The height of the QD is $2.4$~nm and its base diameter  is $18$~nm.
The QDs are located on a $0.6$~nm (one lattice constant) thick InAs wetting
layer. The overall QDs symmetry ($C_{2v}$) results from the combined
zinc-blende lattice symmetry and the QD cylindrically symmetric lens
shape.~\cite{bester-cylindrical,bester} In effect, despite a nominally
cylindrical base, the bright and dark excitonic doublets are
non-degenerate,\cite{bester,zielinski-shape}
%with pronounced bright exciton (BE) ($85\mu$eV) and dark exciton
%(DE)($3.8$~$\mu$eV) splitting.
with pronounced bright exciton (BE) and dark exciton (DE) splitting.  The
bright exciton lines are polarized along non-equivalent $[110]$ and
$[1\underline{1}0]$ crystal axes, with small ($3\%$ to $4\%$) polarization
anisotropy. In both QDs, a fraction of a meV below the lower energy BE spectral
line, a weak emission line that originates from radiative recombination of the DE
is observed.
%The non-zero, out-of-plane polarized emission of the dark exciton has been
%recently suggested~\cite{karlsson1,karlsson2} by group-theoretical approach
%and confirmed by both atomistic~\cite{dark-marek} and experimental
%%studies.~\cite{dark-smolenski}
The visible DE line is polarized along the growth axis ($[001]$) of the QD, and
the intensity of this line is about seven [six] orders of magnitude weaker
than that of the BE in the short (Fig.~\ref{lens}~(a)) [tall
(Fig.~\ref{lens}~(c))] QD. For both QDs, the lowest energy DE
eigenstate remains completely optically inactive, and therefore, it is not shown
in Fig.~\ref{lens}.

We note that we also studied numerous $C_{2v}$ QDs with base shapes varying
from ellipsoidal to a square base of a truncated pyramid (not
shown).~\cite{michler} In these cases, the $z$-polarized dark exciton line can
reach (truncated pyramid) a substantial fraction ($\approx10^{-3}$) of the
bright exciton magnitude, \emph{i.e.}~three to four orders of magnitude stronger
that of the cylindrically symmetric lens-shaped QDs. This increase of dark
exciton activity can be attributed to the increased hole subband
mixing.~\cite{dark-marek} Nevertheless, the emission from the visible DE
eigenstate of $C_{2v}$ QDs is, as expected from symmetry considerations, always
polarized along the QD growth axis.

\section{Low-symmetry quantum dots}

\begin{figure}
	\begin{centering}
		\includegraphics[width=0.5\textwidth]{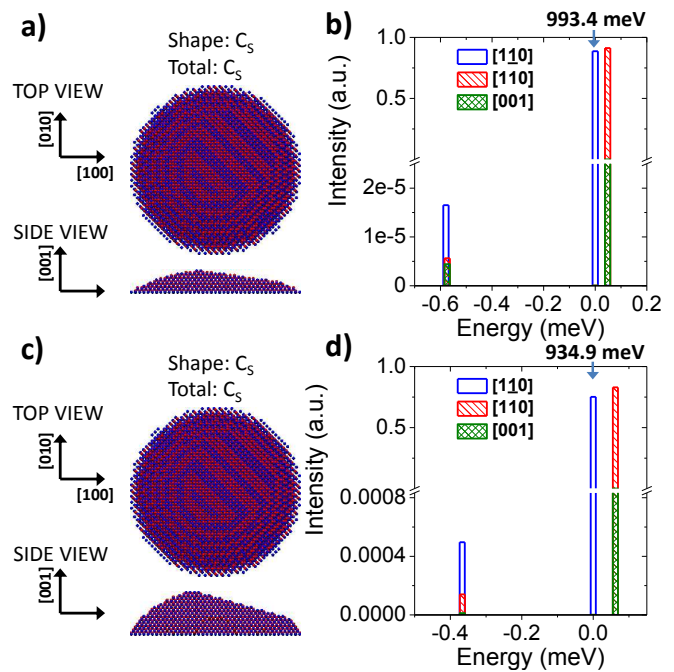}
		\protect
		\caption{ (Color online) (a) {[}(c){]} Schematic description of the lens-shaped
			QD in Fig.~\ref{lens}(a) {[}Fig.~\ref{lens}(c){]} with a truncated
			planar facet. The calculated excitonic spectrum for the QDs in (a)
			and (c) are displayed in (b) and (d) respectively.}
		\label{brokenlens}
		\par
	\end{centering}
\end{figure}

In reality, ideally symmetrized systems of macroscopic scale are extremely
rare. Recent theoretical studies of epitaxial growth of strained
heterostructures~\cite{asymmetric-islands} concluded that self-assembled QDs
can actually grow highly asymmetrically, largely deviating from any rotational
symmetry. Realistic self-assembled QDs thus have symmetries which can deviate
quite substantially from the idealized shapes of circular or ellipsoidal lenses.
% or a truncated pyramid.
In order to methodically study the effects
of the symmetry reduction on the optical properties of the QD, we introduced an
inclined planar facet between the QD and the covering host material. In
Fig.~\ref{brokenlens} (a) and (c), we present schematic depictions of two
modified lens-shaped QDs, where the symmetry has been reduced by truncating
several atomic layers, as depicted in Fig.~\ref{brokenlens}.
Fig.~\ref{brokenlens} (a) and (c) describe the structural modifications made to
the symmetrical QDs in Fig.~\ref{lens} (a) and (c), respectively, while the
calculated excitonic spectra for these QDs is presented in
Fig.~\ref{brokenlens} (b) and (d).

The overall QDs symmetry (structure and lattice) is now $C_{s}$ (reflection
about one perpendicular plane). The choice of this shape-symmetry breaking
mechanism is not intended to reproduce reality, but rather to introduce a
certain well-controlled perturbation to otherwise perfectly circularly
symmetric lens-shaped QDs. The truncation removes only $2\%$ ($8\%$ in (c)) of
the QD atoms, replacing them by the host material atoms. Therefore, the change
in the BE emission energy of $5$~meV ($3$~meV in (d)) is rather small.
Likewise, the BE emission intensities differ by no more than $3\%$ ($6\%$) from
those of the perfect lens in Fig.~\ref{lens} (a) (Fig.~\ref{lens} (c)). Both BE
lines are in-plane polarized along the $[110]$ and $[1\underline{1}0]$
crystallographic directions. Consistently, the BE-DE energy difference of
$0.57$ meV ($0.37$ meV) is comparable with the $0.515$ meV ($0.315$ meV) of the
symmetrical QD. The BE fine structure splitting (FSS) is $48~\mu$eV
($28~\mu$eV), notably different from the $85~\mu$eV ($41~\mu$eV) observed for
the symmetrical lens-shaped QD. This should not come as a surprise since the
BE FSS is very sensitive to the QD structural shape.\cite{bester,bayer} The
calculated energies are summarized in Table~\ref{tab:summary}.
%The calculated DE splitting is $5~\mu$eV in Fig.~\ref{brokenlens} (b) and
%considerably smaller, $0.5~\mu$eV, in Fig.~\ref{brokenlens} (d).

Yet the most notable difference in the spectra from the asymmetric QDs is that
the DE reveals in-plane linearly polarized emission. The intensity of the much
brighter DE spectral line reaches $10^{-5}$ and $5\times10^{-4}$ of the BE
emission intensity for the QD in (a) and in (c), respectively. The polarization
directions follow those of the BEs, with the brighter one polarized like the
lower energy BE line along the [1$\underline{1}$0] crystallographic direction.
Apparently, both DE eigenstates gain some in-plane polarized oscillator
strength, which is larger than their out of plane polarizations.  The spectrum
presented in Fig.~\ref{brokenlens} (d), in which the dipole moment of the much
brighter DE eigenstate is about $1/1500$ that of the BEs, is in quantitative
agreement with recently measured experimental results.~\cite{dark-experiment}

In the following, we discuss in more detail the effect of structural symmetry
reduction on the exciton emission spectrum. We note first that, in atomistic
calculations, due to the inclusion of the spin-orbit interactions on every atom
and the presence of the underlying crystal lattice, neither the spin nor the
orbital angular momentum are good quantum numbers, even for a perfectly
circularly symmetric lens-shaped QD. Therefore, neither the calculated electron
nor the hole ground states are eigenfunctions of the spin or the total angular
momentum operators. Yet, in the absence of time reversal breaking interactions
such as an external magnetic field, both the electron and the hole states are
doubly (Kramers') degenerate. We label these four states as $e_{1}$, $e_{2}$
and $h_{1}$, $h_{2}$, respectively. The exciton energies are then calculated by
including the Coulomb interaction between the electron and hole ground states.
The direct Coulomb attraction reduces the energy of the exciton, while the
exchange interactions remove the degeneracy between the four possible exciton
eigenfunctions.~\cite{bayer,ivchenko,takagahara}
%We note here that the actual excitonic eigen-energies are further affected by
%relatively small corrections due to mixing with higher-lying shells.

Using the exciton states
($\ket{\mathrm{BE1}} =\ket{e_{2}h_{2}} $,
$\ket{\mathrm{BE2}} =\ket{e_{1}h_{1}} $,
$\ket{\mathrm{DE1}} =\ket{e_{1}h_{2}} $,
$\ket{\mathrm{DE2}} =\ket{e_{2}h_{1}} $)
as a basis,
the following exchange Hamiltonian is obtained for a system with $C_{2v}$
symmetry. This general shape of the exchange Hamiltonian is anticipated
from basic symmetry considerations.~\cite{bayer,ivchenko,takagahara}
We use our atomistic model and CI calculations to extract the matrix
elements for this Hamiltonian,
\begin{equation}
	\mathcal{H}_{C_{2v}}=\frac{1}{2}
	\left[\begin{array}{cccc}
			\Delta_{0}                    & \mathrm{e^{-i\pi/2}}\Delta_{1} & 0           & 0\\
			\mathrm{e^{i\pi/2}}\Delta_{1} & \Delta_{0}                     & 0           & 0\\
			0                             & 0                              & -\Delta_{0} &  \Delta_{2}\\
			0                             & 0                              &  \Delta_{2} & -\Delta_{0}
	\end{array}\right]
	\label{hamc2v}
\end{equation}
where $\Delta_{1}$ and $\Delta_{2}$ describe the DE and BE energy splitting,
respectively, and $\Delta_{0}$ corresponds to the energy difference between the
DE and the BE exciton doublets. The phase factors for the interaction between
the BE configurations $\ket{\mathrm{BE1}} $ and $\ket{\mathrm{BE2}} $
are chosen so that their dipole moments are polarized parallel and
perpendicular to the $[110]$ crystallographic direction, and $\Delta_{1}$ and
$\Delta_{2}$ remain real numbers.~\cite{garnett-prl,ivchenko} Normally, and in
particular for nanosystems with a non-circularly symmetric base,
%(such as ellipsoidal or pyramidal QDs),
the excitonic basis should be extended to include higher than ground electron
and hole states, yielding a higher-dimensional
Hamiltonian.\cite{zielinski-shape} Nevertheless, the $4\times4$ Hamiltonian
qualitatively captures most of the physical properties of the QD-confined
exciton fine structure.  In the results reported here, the actual corrections to
the calculated excitonic eigenenergies are rather small.

As seen from the Hamiltonian for the symmetric QD (Eq.~\ref{hamc2v}), the DE
and BE subspaces do not mix. The weak optical activity of one of the DE
eigenstates is due to small, non-zero $\hat{z}$-polarized dipole matrix elements
$\left| \braket{e_{1}|\hat{z}|h_{2}} \right|
=\left| \braket{e_{2}|\hat{z}|h_{1}} \right|$,
attributed predominantly to light-hole/heavy-hole mixing. These
contributions interfere constructively (destructively) for the upper (lower)
energy DE eigenstate.  The in-plane polarized ($\hat{x}$ and $\hat{y}$) DE
dipole moments for the QD of $C_{2v}$ symmetry vanish:
$\left| \braket{e_{1}|\hat{x},\hat{y}|h_{2}} \right|
=\left| \braket{e_{2}|\hat{x},\hat{y}|h_{1}} \right| = 0$.

The Hamiltonian (Eq.~\ref{hamc2v}) is expressed in the basis of the four
excitonic states calculated using the atomistic, tight-binding model. In the
atomistic calculation, neither single particle states nor the excitonic
configurations are eigenfunctions of the total angular momentum. In the
literature,~\cite{ivchenko,bayer} however, the electron-hole exchange
Hamiltonian is customarily expanded in the basis of the excitons with total
angular momentum projections on the QD symmetry axis:
$\ket{\mathrm{+1}}$,
$\ket{\mathrm{-1}}$,
$\ket{\mathrm{+2}}$,
$\ket{\mathrm{-2}}$.
In this basis, the states with $M=\pm2$ cannot couple to light and they are
termed DEs. The only optically active states are $M=\pm1$ and they are termed
BEs. Since, as discussed above, the the $4\times4$ Hamiltonian in both cases has
exactly the same formal structure, one can readily make a one-to-one correspondence
between the basis elements  of both models and use the atomistic model to
calculate the parameters of the Hamiltonian in the angular momentum
representation.

For the QDs with lower symmetry ($C_{s}$), the in-plane dipole
moments do not vanish, \emph{i.e.}~$\left| \braket{e_{1}|\hat{x},\hat{y}|h_{2}} \right| \neq0$,
allowing, in principle, DE emission (and absorption) along the growth
direction of the QD. Depending on the deviation from symmetry, the
magnitude of $\left|\braket{e_{1}|\hat{x}|h_{2}} \right|$
can reach a substantial fraction (up to $3\%$) of that of the BE, $\left| \braket{e_{1}|\hat{x}|h_{1}} \right|$.

The electron-hole exchange Hamiltonian for the low-symmetry QD ($C_{s}$),
as calculated by the atomistic CI model, is
%*****SLAVA******
\begin{equation}
	\mathcal{H}_{C_{s}}=\frac{1}{2}
		\left[\begin{array}{cccc}
				\Delta_{0}                     & \mathrm{e^{-i\pi/2}}\Delta_{1}   & \mathrm{e^{-i\pi/4}}\Delta_{11} & \mathrm{e^{-i\pi/4}}\Delta_{12}\\
				\mathrm{e^{i\pi/2}}\Delta_{1}  & \Delta_{0}                       & \mathrm{-e^{i\pi/4}}\Delta_{12} & \mathrm{-e^{i\pi/4}}\Delta_{11}\\
				\mathrm{e^{i\pi/4}}\Delta_{11} & \mathrm{-e^{-i\pi/4}}\Delta_{12} & -\Delta_{0}                     &  \Delta_{2}\\
				\mathrm{e^{i\pi/4}}\Delta_{12} & \mathrm{-e^{-i\pi/4}}  \Delta_{11} &  \Delta_{2}                     & -\Delta_{0}
	\end{array}\right]
	\label{hamcs}
\end{equation}

The main difference between this and the $C_{2v}$ Hamiltonian (Eq.~\ref{hamc2v})
are the non-vanishing DE-BE mixing terms. The DE-BE mixing is
determined by two matrix elements  $\Delta_{11}$ and $\Delta_{12}$, with a
phase. The phase reflects the $\pi/4$ phase associated with the BE eigenstates
and the zero phase associated with the DE eigenstates of Eq.~\ref{hamc2v}.
Thus, there are two sources for the DE dipole moment. The first is the non-zero
$\hat{z}$-polarized dipole moment discussed above, and the second is the
in-plane polarized dipole moments resulting from mixing with the BEs. Neither
of these effects can be \emph{a priori} neglected.

The formal structure of the $C_{s}$ Hamiltonian (Eq.~\ref{hamcs}) resembles the
Hamiltonian for a QD in an external magnetic field in the Voigt
geometry.~\cite{bayer} One can therefore view the mixing terms as resulting
from an effective in-plane magnetic field. An intuitive way of seeing the
source for this magnetic field is to describe the hole motion in the magnetic
field produced by the spinning electron and vice versa. The spinning electron
thus couples the $\ket{\mathrm{DE1}}$ ($\ket{\mathrm{DE2}}$) exciton to the
$\ket{\mathrm{BE2}}$ ($\ket{\mathrm{BE1}}$) exciton, resulting in the mixing term $\Delta_{12}$, while the spinning hole
couples the $\ket{\mathrm{DE1}}$ ($\ket{\mathrm{DE2}}$) to the
$\ket{\mathrm{BE1}}$ ($\ket{\mathrm{BE2}}$) exciton, resulting in the mixing term $\Delta_{11}$. The effective in-plane
magnetic field results from the fact that, in the lower symmetry QD, both the
electron and the hole have non-vanishing in-plane spin projection expectation
values. For the QD in Fig. \ref{brokenlens}, the facet effectively tilts the QD
axis with respect to the crystallographic direction $[001]$.  This
``tilt'' results in non-zero expectation values for the in-plane
spin-components of both the electron and the hole.
In contrast, for a QD with $C_{2v}$ symmetry these expectation values
are exactly zero.

%The Hamiltonian from Eq.~\ref{hamcs} was expanded in the basis of four excitonic states calculated using the tight-binding model.
%In the atomistic calculation, neither single particle states nor the excitonic configurations are eigenfunctions of the %total angular momentum.
%In the literature,~\cite{ivchenko,bayer} however, the exchange matrix is customarily expanded in the basis of the total %angular momentum projections:
%$\ket{\mathrm{+1}}$,
%$\ket{\mathrm{-1}}$,
%$\ket{\mathrm{+2}}$,
%$\ket{\mathrm{-2}}$.
%In this language, the states with $M=\pm2$ cannot couple to light and the only optically active states are $M=\pm1$.

Since there is a one-to-one correspondence between the atomistic model basis and
the angular momentum basis of Eq.~\ref{hamc2v}, the exchange Hamiltonian of
Eq.~\ref{hamcs} must have the same formal structure in the angular momentum
basis as well. Therefore, we can fit the atomistic model calculated
excitonic spectra for obtaining the mixing parameters $\Delta_{11}$ and
$\Delta_{12}$ of the simplified $4\times4$ Hamiltonian expressed by the angular
momentum basis of the symmetric QD as in Eq.~\ref{hamc2v}.

The calculated and fitted values of all the entries in the Hamiltonians
(Eq.~\ref{hamc2v} and Eq.~\ref{hamcs}) are depicted in Table~\ref{tab:summary}.
With these entries, the eigenenergies and eigenstates of the four excitons can
be straightforwardly calculated.  One immediately notes that, in all cases, the
symmetric (under electron-hole exchange) BE state couples only to the
antisymmetric DE state and the antisymmetric BE state couples only to the
symmetric DE state. Thus, the visible DE state has opposite symmetry to the
symmetry of the BE state to which it couples (the one polarized along the
$[1\underline{1}0]$ axis). Indeed, the visible DE state of the QD in
Fig.~\ref{brokenlens} (c) has even symmetry while the $[1\underline{1}0$ polarized BE exciton
to which it couples has odd symmetry under electron-hole exchange.
\begin{table*}
	\begin{centering}
		\protect\caption{Summary of the calculated values in the presented systems\label{tab:summary}}
		\smallskip{}
		\begin{tabular}{ccccccl}
			\toprule
%			Parameter       & Units     & \multicolumn{4}{c}{Value}                                                                         & Description       \\
%			\midrule
			~               &           & Fig.~\ref{lens} (a) & Fig.~\ref{lens} (c) & Fig.~\ref{brokenlens} (a) & Fig.~\ref{brokenlens} (c) &                   \\
			~               &           & (flat, symmetric)  & (tall, symmetric)    & (flat, truncated)         & (tall, truncated)         &                   \\
			\midrule
			$\Delta_{0}$    & $\mu$eV   & 515                 & 315                 & 570                       & 370                       & BE-DE split       \\
			$\Delta_{1}$    & $\mu$eV   & 84.7                & 40.7                & 48                        & 28                        & BE FSS amplitude  \\
			$\Delta_{2}$    & $\mu$eV   & 3.88                & -6.51               & -5.0                      & -0.5                      & DE FSS amplitude  \\
			\midrule
			$\Delta_{11}$   & $\mu$eV   & 0                   & 0                   & 3.71                      & 3.27                      & BE-DE mixing      \\
			$\Delta_{12}$   & $\mu$eV   & 0                   & 0                   & 0.81                      & 12.7                      & BE-DE mixing      \\
			\bottomrule
		\end{tabular}
		\par
	\end{centering}
\end{table*}

The apparent anisotropy between the dipole moments of the two DE eigenstates
can be traced to the constructive contribution, due to DE-BE mixing, to the
optical activity of one DE eigenstate and destructive contribution to the
optical activity of the other eigenstate. If the magnitudes of $\Delta_{11}$
and $\Delta_{12}$ are equal, the destruction is then complete, leading again 
to only one (like for symmetrical QDs) optically active DE eigenstate. 
Here, however, the visible DE is polarized in-plane.
%
%$\left| \braket{ e_{1}|\hat{s}_{x}|e_{1} } \right|
%=\left| \braket{ e_{1}|\hat{s}_{y}|e_{1} } \right|
%=\left| \braket{ h_{1}|\hat{s}_{x}|h_{1} } \right|
%=\left| \braket{ h_{1}|\hat{s}_{y}|h_{1} } \right|=0$.
% \emph{e.g.} $\left|\braket{h_{1}|\hat{s}_{x}|h_{1}}\right| \approx 0.018 \neq 0$.

% ======= MICHAL SUPPLY THE ELECTRON OR REMOVE THIS INFORMATION ====== %
%In effect, this is an intuitive explanation for the BE-DE mixing that results
%from the QD lower symmetry and the non-zero in-plane dipole moment mainly for
%only one DE eigenstate.  Finally, in Fig.~\ref{brokenlens2}, we present a
%system with larger structural asymmetry as compared to the circularly
%symmetric lens-shaped QD.  This QD is obtained by truncating a lens-shaped QD
%of the same base diameter as the lens discussed earlier, but with twice the
%height ($4.8$~nm). The QD is truncated by a steeper facet, which reduces the
%QD height after truncation to $4.2$~nm.  This more deformed QD results in
%larger dipole moment and more pronounced DE polarization anisotropy.  For the
%structure depicted in in Fig.~\ref{brokenlens} (c), the DE in-plane dipole
%moment magnitude is about $1/1500$ of the dipole moment of the lower energy
%eigenstate of the BE and it is polarized in the same in plane direction
%{[}1$\underline{1}$0{]}, in quantitative agreement with the data reported
%recently.~\cite{dark-experiment}

We carefully note here that the lower-symmetry QDs presented above can be
viewed as a proof of concept only. Though clearly most real QDs have lower than
$C_{2v}$ symmetry, a reliable comparison with experiment is impossible without
a detailed knowledge of the QD structural shape and its composition and strain
distributions. Systematic comparison with experimental measurements must take
into account as well the mixed composition (In$_{x}$Ga$_{1-x}$As) of the QD and
the resulting lattice disorder. While such studies are beyond the scope of this
manuscript, we note that if one assumes a uniform composition profile, the
conclusions discussed above remain qualitatively correct for In$_{x}$Ga$_{1-x}$As
disordered systems.

\section{Conclusions \label{sec:conclusions}}

We use atomistic model simulations for studying the effect of symmetry reduction
on the optical properties of excitons confined in self-assembled InAs/GaAs QDs.
We compare between $C_{2v}$ symmetrical QDs and asymmetrical ones with $C_{s}$
symmetry only. The symmetry in the studied QD is reduced by forming a planar
facet between the QD and its host material. We show that, while the symmetry
reduction hardly affects the BE eigentates' oscillator strength and
polarizations, it strongly affects the DE eigenstates. For symmetric QDs, one DE
eigenstate is dark while the other one has a small dipole moment polarized along
the QD symmetry axis. For the symmetry ``broken'' QDs,
the DE eigenstates are mixed with the BE eigenstates of opposite symmetry under
electron-hole exchange. Therefore, they become optically active for in-plane
polarized light, like the BEs with which they are mixed. There is very large
anisotropy between the dipole moments of the two DE eigenstates, since there
are two mixing terms that interfere constructively for one DE eigenstate and
destructively for the other one.  The calculated excitonic spectrum and its
polarization selection rules are in quantitative agreement with recently
obtained experimental results.

~

\section{Acknowledgements}

We acknowledge support from the Polish National Science Centre based
on decision DEC-2011/01/D/ST3/03415. Support from the Israeli Science
Foundation, and from the Technion's RBNI is acknowledged as well.
Stimulating discussions with G.~W.~Bryant are acknowledged.

%%%%%%%%%%%%%%%%%%%%%%%%%%%%%% Backmatter

\bibliographystyle{apsrev4-1}
\bibliography{dark_exciton_bib}

\end{document}